\documentclass[10pt,a4paper]{article}

\usepackage{times}

\usepackage{epsfig,wrapfig}
\usepackage{epstopdf} % for pdflatex

\usepackage{fancyhdr}

\usepackage[lmargin=2.5cm, rmargin=2.5cm,tmargin=3.50cm,bmargin=3.50cm]{geometry}
\usepackage{indentfirst}

\usepackage{titlesec}
\titleformat{\section}[hang]
  {\centering}{\thesection}{1ex}{\normalsize \textsc}%%
\titleformat{\subsection}[hang]
  {}{\thesubsection}{1ex}{\normalsize \textit}%%

\renewcommand{\thesection}{ \normalsize \textnormal{\Roman{section}.}}
\renewcommand{\thesubsection}{\normalsize \textnormal{\textsc{\textit{\Alph{subsection}.}}}}

\def\e{\begin{equation}}
\def\f{\end{equation}}
\def\_#1{{\bf #1}}

\def\.{\cdot}

\usepackage{bm}
\usepackage[usenames,dvipsnames]{xcolor}
\usepackage{srcltx}
\usepackage{psfrag}
\usepackage{epsfig}
\usepackage{graphicx}
\usepackage{epstopdf}
\usepackage{color}
\usepackage[tbtags,sumlimits,nointlimits,reqno]{amsmath}
\usepackage{amssymb}
\usepackage{color}
\usepackage{float}
\usepackage{subfigure}
\usepackage{gensymb}
\usepackage[mode=errorstop]{pstool}
\begin{document}

%%% Title of paper
\title{\large \textbf{Angle-Independent Nongyrotropic Metasurfaces}}
%
%%% Author(s) and affiliation
\def\affil#1{\begin{itemize} \item[] #1 \end{itemize}}
\author{\normalsize \bfseries \underline{G. Lavigne}$^1$ and C. Caloz$^1$}
\date{}
\maketitle
\thispagestyle{fancy} % header also to the first page
\vspace{-6ex}
\affil{\begin{center}\normalsize $^1$Polytechnique Montr\'eal, Department of Electrical Engineering, Blvd. Edouard-Montpetit, H3T 1J4, Montr\'eal, Canada  \\
guillaume.lavigne@polymtl.ca
 \end{center}}

%%% Abstract
\begin{abstract}
\noindent \normalsize
\textbf{\textit{Abstract} \ \ -- \ \
We derive a general condition for angle-independent bianisotropic nongyrotropic metasurfaces and present two applications corresponding to particular cases: an angle-independent absorber/amplifier and an angle-independent spatial gyrator.
}
\end{abstract}

\section{Introduction}\label{sec:intro}

Metasurfaces are generally designed for a given specification, namely the incident, reflected and transmitted fields with specific angles, frequencies and polarizations~\cite{glybovski2016metasurfaces,achouri2018design,asadchy2018bianisotropic}. Specifically, general bianisotropic metasurfaces have been shown to have intricate and diverse angular scattering properties~\cite{achouri2018influence}. It would be particularly desirable for many applications to devise metasurfaces that would exhibit the same response for different excitations.

Here, we theoretically derive bianisotropic nongyrotropic metasurfaces whose response is angle-independent. Based on this result, we study two potential applications of angle-independent bianisotropic metasurafaces: an absorber/amplifier and a spatial gyrator.

\section{Theoretical Derivation}

We consider the problem of an uniform bianisotropic metasurface surrounded by air placed in the $xy$ plane $z=0$ . Such a metasurface can be efficiently modelled as a zero-thickness discontinuity in space using generalized sheet transition conditions~(GSTCs) and surface susceptibility tensors~\cite{achouri2014general}. The tangential susceptiblity-based GSTCs model of such a metasurface is

\begin{subequations}\label{eq:GSTC}
\begin{equation}
\hat{z} \times \Delta\mathbf{H} = j \omega \epsilon \overline{\overline{ \chi}}_\text{ee} \mathbf{E}_\text{av} +  j k \overline{\overline{ \chi}}_\text{em}   \mathbf{H}_\text{av} ,
\end{equation}
\begin{equation}
\Delta \mathbf{E} \times \hat{z}   = j k \overline{\overline{ \chi}}_\text{me}  \mathbf{E}_\text{av} + j \omega \mu \overline{\overline{ \chi}}_\text{mm} \mathbf{H}_\text{av}.
\end{equation}
\end{subequations}
where the symbol $\Delta$ and the subscript 'av' represent the differences and averages of the tangential electric or magnetic fields at both sides of the metasurface, and $\overline{\overline{ \chi}}_\text{ee}$, $\overline{\overline{ \chi}}_\text{mm}$, $\overline{\overline{ \chi}}_\text{em}$, $\overline{\overline{ \chi}}_\text{me}$ are the bianisotropic susceptibility tensors characterizing the metasurface.

We seek here a general condition for metasurface angle-independent transformation, with the only restriction of nongyrotropy to avoid excessive complexity at this point. For the sake of conciseness, we limit the investigation to the $s$ polarization, and a similar treatment naturally applies to the $p$ polarization. Under such conditions, Eqs.~\eqref{eq:GSTC} reduce to

\begin{subequations}\label{eq:GSTC_TE}
\begin{equation}
\Delta H_x = j \omega \epsilon \chi_\text{ee}^{yy} E_{y,\text{av}} + j k \chi_\text{em}^{yx} H_{x,\text{av}},
\end{equation}
\begin{equation}
\Delta E_y = j k \chi_\text{me}^{xy} E_{y,\text{av}} + j \omega \mu \chi_\text{mm}^{xx} H_{x,\text{av}},
\end{equation}
\end{subequations}
where the field differences and averages are related to the forward-incidence transmission and reflection coefficients ($T_1$ and $R_1$) as

\begin{subequations}\label{eq:av_diff_T_R}
\begin{equation}
\Delta H_x = (T_1-(1+R_1))H_{x,\text{i}}, \quad H_{x,\text{av}} = \frac{1}{2} (T_1+R_1+1)H_{x,\text{i}},
\end{equation}
\begin{equation}
\Delta E_y = (T_1-(1-R_1))E_{y,\text{i}}, \quad E_{y,\text{av}} = \frac{1}{2}(T_1+1-R_1)E_{y,\text{i}}.
\end{equation}
%\begin{equation}
%H_{x,\text{av}} = \frac{1}{2} (T_1+R_1+1)H_{x,\text{i}},
%\end{equation}
%\begin{equation}
%E_{y,\text{av}} = \frac{1}{2}(T_1+1-R_1)E_{y,\text{i}}.
%\end{equation}
\end{subequations}
The ratio of the tangential electric and magnetic fields in~\eqref{eq:av_diff_T_R} as a function of the incidence angle $\theta$ can be written as
\begin{equation}\label{eq:ratio_Ey_Hx}
\frac{E_{y,\text{i}}}{H_{x,\text{i}}} = \frac{\eta_0}{\cos \theta},
\end{equation}
where $\eta_0$ is the wave impedance in free space.
Inserting~\eqref{eq:av_diff_T_R} into~\eqref{eq:GSTC_TE}, respectively dividing the resulting equations by $H_{x,\text{i}}$ and by $E_{y,\text{i}}$ and substituting~\eqref{eq:ratio_Ey_Hx} yields the following expression for the forward transmission and reflection coefficients in terms of the susceptibilities:
\begin{subequations}\label{eq:T1_R1}
\begin{equation}
T_1 = \frac{(2+j k \chi_\text{em}^{yx})(-2j +k \chi_\text{me}^{xy})- j \epsilon \mu \omega^2 \chi_\text{ee}^{yy} \chi_\text{mm}^{xx}}{j \frac{\eta}{\cos \theta} 2 \epsilon \chi_\text{ee}^{yy} \omega + 2 \mu \chi_\text{mm}^{xx} \omega \frac{ \cos \theta}{\eta} -j( 4 + k^2 \chi_\text{em}^{yx} \chi_\text{me}^{xy} - \epsilon \mu \omega^2 \chi_\text{ee}^{yy} \chi_\text{mm}^{xx})},
\end{equation}
\begin{equation}
R_1 = \frac{2(k(\chi_\text{me}^{xy}- \chi_\text{em}^{yx})+\frac{\eta}{\cos \theta}  \epsilon \chi_\text{ee}^{yy} \omega-\mu \chi_\text{mm}^{xx} \omega \frac{ \cos \theta}{\eta})}{j \frac{\eta}{\cos \theta} 2 \epsilon \chi_\text{ee}^{yy} \omega + 2 \mu \chi_\text{mm}^{xx} \omega \frac{ \cos \theta}{\eta} -j( 4 + k^2 \chi_\text{em}^{yx} \chi_\text{me}^{xy} - \epsilon \mu \omega^2 \chi_\text{ee}^{yy} \chi_\text{mm}^{xx})}.
\end{equation}
\end{subequations}
Following the same procedure for the backward case yields
\begin{subequations}\label{eq:T2_R2}
\begin{equation}
T_2 =\frac{(-2+ j k \chi_\text{em}^{yx})(2j +k \chi_\text{me}^{xy})- j \epsilon \mu \omega^2 \chi_\text{ee}^{yy} \chi_\text{mm}^{xx}}{j \frac{\eta}{\cos \theta} 2 \epsilon \chi_\text{ee}^{yy} \omega + 2 \mu \chi_\text{mm}^{xx} \omega \frac{ \cos \theta}{\eta} -j( 4 + k^2 \chi_\text{em}^{yx} \chi_\text{me}^{xy} - \epsilon \mu \omega^2 \chi_\text{ee}^{yy} \chi_\text{mm}^{xx})},
\end{equation}
\begin{equation}
R_2 = \frac{2(k( \chi_\text{em}^{yx}-\chi_\text{me}^{xy})+\frac{\eta}{\cos \theta}  \epsilon \chi_\text{ee}^{yy} \omega-\mu \chi_\text{mm}^{xx} \omega \frac{ \cos \theta}{\eta})}{j \frac{\eta}{\cos \theta} 2 \epsilon \chi_\text{ee}^{yy} \omega + 2 \mu \chi_\text{mm}^{xx} \omega \frac{ \cos \theta}{\eta} -j( 4 + k^2 \chi_\text{em}^{yx} \chi_\text{me}^{xy} - \epsilon \mu \omega^2 \chi_\text{ee}^{yy} \chi_\text{mm}^{xx})}.
\end{equation}
\end{subequations}

Inspecting~\eqref{eq:T1_R1} and~\eqref{eq:T2_R2} reveals that a metasurface transformation generally has an angle dependance that is associated with the $\cos \theta$ and $\frac{1}{\cos \theta}$ coefficients in the above expressions of the reflection and transmission coefficients. However, these coefficients only affect the $\chi_\text{ee}^{yy}$ and $\chi_\text{mm}^{xx}$ susceptibilities. Therefore, imposing the restriction $\chi_\text{ee}^{yy}=\chi_\text{mm}^{xx} =0$ removes the angular dependance. Practically, even if the $\chi_\text{ee}^{yy}$ and $\chi_\text{mm}^{xx}$ susceptiblities are not exactly zero but still negligeable compared to the $\chi_\text{em}^{yx}$ and $\chi_\text{me}^{xy}$ susceptiblities, the response of the metasurface will be essentially independent of the angle, except towards grazing angles $\frac{1}{\cos \theta} \rightarrow \infty$. Inserting this condition into~\eqref{eq:T1_R1} and~\eqref{eq:T2_R2} yields then the angle-independent scattering coefficients

\begin{subequations}\label{eq:T_R_simplified}
\begin{equation}
T_1 = \frac{(2j- k \chi_\text{em}^{yx})(-2j +k \chi_\text{me}^{xy})}{  4 + k^2 \chi_\text{em}^{yx} \chi_\text{me}^{xy}}, \quad R_1 = \frac{2 j k(\chi_\text{me}^{xy}- \chi_\text{em}^{yx})}{ 4 + k^2 \chi_\text{em}^{yx} \chi_\text{me}^{xy}},
\end{equation}
\begin{equation}
T_2 =\frac{(-2j- k \chi_\text{em}^{yx})(2j +k \chi_\text{me}^{xy})}{ 4 + k^2 \chi_\text{em}^{yx} \chi_\text{me}^{xy}}, \quad R_2 =\frac{2 j k( \chi_\text{em}^{yx}-\chi_\text{me}^{xy})}{ 4 + k^2 \chi_\text{em}^{yx} \chi_\text{me}^{xy}}.
\end{equation}
\end{subequations}

\section{Angle-Independent Absorber/Amplifier}

An extremely useful application of such an angle-independent metasurface would be that of an angle-independent absorber\footnote{A bianisotropic metasurface was demonstrated as thin-absorber was theoretical and experimentally demonstrated in [6] and [7], respectively. However, this metasurface absorber is not angle-independent.}.

Inspecting~\eqref{eq:T_R_simplified} shows that the reflection at both sides of the metasurface is suppressed by imposing $\chi_\text{em}^{yx} = \chi_\text{me}^{xy}$ since this leads to $R_1 = R_2 = 0$. Note that such a condition implies nonreciprocity since as it violates the reciprocity condition $\overline{\overline{\chi}}_\text{em} = -\overline{\overline{\chi}}_\text{me}^T$, which reveals that nonreciprocity is a fundamental condition for metasurface angle-independent absorption. The corresponding foward and backward transmission coefficients reduce then to

\begin{subequations}\label{eq:T_simplified}
\begin{equation}\label{eq:T_simplified_a}
T_1 = \frac{-k^2 (\chi_\text{em}^{yx})^2+4 j k \chi_\text{em}^{yx}+4}{  4 + k^2 (\chi_\text{em}^{yx})^2},
\end{equation}
\begin{equation}\label{eq:T_simplified_b}
T_2 =\frac{-k^2 (\chi_\text{em}^{yx})^2-4 j k \chi_\text{em}^{yx}+4}{ 4 + k^2 (\chi_\text{em}^{yx})^2}.
\end{equation}
\end{subequations}

Equations~\eqref{eq:T_simplified} are analysis equations giving the transmission coefficients through a metasurface of susceptibility $\chi_\text{em}^{yx}$. For design, we need an inverse, synthesis equation. Such an equations is obtained by solving~\eqref{eq:T_simplified_a} for a specific $T_1$, $T_{1,\text{spec}}$, which yields

\begin{equation}\label{eq:chi_from_T1}
\chi_\text{me}^{xy} = \chi_\text{em}^{yx} = \frac{2i}{k}\frac{1- T_{1,\text{spec}}}{1+ T_{1,\text{spec}}},
\end{equation}

whose insertion in\eqref{eq:T_simplified_b} yields

\begin{equation}\label{eq:T2_from_T1}
T_2 = \frac{1}{T_{1,\text{spec}}}.
\end{equation}

Equation~\eqref{eq:T2_from_T1} reveals that specifying $T_1$ via~\eqref{eq:chi_from_T1} results in having $T_2$ being the opposite of this $T_1$. This physically means that an angle-independent absorption corresponding to $T_{1,\text{spec}}$ in the forward excitation implies a gain of the same level in the opposite direction. Therefore, the overall metasurface needs to be active. Such a metasurface may be realized by integrating transistors in the metasurface~\cite{kodera2011artificial}. The topic of transistor-based metasurface absorber was previously discussed in~\cite{li2017high}.

\section{Angle-Independent Spatial Gyrator}

Another interesting application would be an angle-independent spatial gyrator. Given the fundamental nature of the gyrator as nonreciprocal component~\cite{tellegen1948gyrator}, one may easily anticipate that an angle-independent spatial could enable many unique devices. Such a gyrator may be realized by specifying zero reflection and $T_1 = 1e^{i \pi/2}$, so that $T_2 = 1e^{-i \pi/2}$ according to~\eqref{eq:T2_from_T1}, which indeed corresponds to a $\pi$ phase difference between the forward and backward transmissions. The corresponding susceptibilities are found from~\eqref{eq:chi_from_T1} as

\begin{equation}\label{eq:chi_gyrator}
\chi_\text{me}^{xy} = \chi_\text{em}^{yx} = \frac{2}{k}.
\end{equation}

\section{Conclusion}\label{sec:conclusion}
We have derived a general condition for an angle-independent nongyrotropic metasurface, and theoretically derived two fundamental applications: an angle-independent absorber/amplifier and an angle-independent spatial gyrator.

%% References

\small{

\bibliography{LIB}
\bibliographystyle{ieeetr}

}

\end{document}